\documentclass[]{emulateapj}
\usepackage{graphicx}
\begin{document}

\title{The Spectral Slope and Kolmogorov Constant of MHD turbulence}
\author{A. Beresnyak}
\affiliation{Los Alamos National Laboratory, Los Alamos, NM, 87545}

\begin{abstract}
  The spectral slope of strong MHD turbulence has recently been a matter of controversy. While
  Goldreich-Sridhar model (1995) predicts Kolmogorov's -5/3 slope of turbulence, shallower slopes
  were often reported by numerical studies. We argue that earlier numerics was affected by driving
  due to a diffuse locality of energy transfer in MHD case.  Our highest-resolution simulation
  ($3072^2\times 1024$) has been able to reach the asymptotic $-5/3$ regime of the energy
  slope. Additionally, we found that so-called dynamic alignment, proposed in the model with -3/2
  slope, saturates and therefore can not affect asymptotic slope.  The observation of the asymptotic
  regime allowed us to measure Kolmogorov constant $C_{KA}=3.2\pm 0.2$ for purely Alfv\'enic turbulence and
  $C_K=4.1\pm 0.3$ for full MHD turbulence. These values are much higher than the hydrodynamic value of
  1.64. The larger value of Kolmogorov constant is an indication of a fairly inefficient energy
  transfer and, as we show in this Letter, is in theoretical agreement with our observation of
  diffuse locality.  We also explain what has been missing in numerical studies that reported
  shallower slopes.
\end{abstract}

\section{Introduction}
The equations of incompressible ideal magnetohydrodynamics, written in terms of Elsasser variables,

\begin{equation}
\partial_t{\bf w^\pm}+\hat S ({\bf w^\mp}\cdot\nabla){\bf w^\pm}=0,\label{mhd}
\end{equation}

where $\hat S= (1-\nabla\Delta^{-1}\nabla)$ is a solenoidal projection, and ${\bf w^\pm}$ (Elsasser
variables) are ${\bf w^+=v+b}$ and ${\bf w^-=v-b}$, ${\bf b=B}/(4\pi \rho)^{1/2}$ are remarkably
similar to the Euler's equation. However, ${\bf w}^\pm$ have different transformation properties than
${\bf v}$.  While in hydrodynamic turbulence the local average velocity can always be excluded by
the choice of reference frame, in MHD ${\bf w}^\pm$ always contain the average local mean
magnetic field that can not be excluded. This leads to a situation when large scale magnetic field
is much stronger than small-scale turbulent perturbations and dynamics is dominated by this local
mean magnetic field \citep{iroshnikov, kraichnan}.  But turbulence does not become weaker down
the cascade as was proposed in aforementioned models.  A proper perturbation theory
\citep{galtier2000} revealed that MHD turbulence has a tendency of becoming stronger on smaller
scales, rather than weaker, due to the fact that the cascade increases perpendicular wavenumber
$k_\perp$, keeping parallel wavenumber $k_\|$ constant.  The ``strength'' of turbulence, as the ratio of
the mean-field term to the nonlinear term can be approximated as $\xi=wk_\perp/v_A k_\|$ and can
increase due to the increasing anisotropy of perturbations $k_\perp/k_\|$ down the cascade.  As
turbulence becomes marginally strong ($\xi\sim 1$), i.e., the linear term is comparable to the
nonlinear term, the cascading timescales become close to the dynamical timescales
$\tau_{casc}\sim\tau_{dyn}=1/wk_\perp$. However, as was argued in \citet{GS95}, the perturbation
frequency $\omega$ has a lower bound due to an uncertainty relation
\footnote{Another bound on $\omega$ that follow from the uncertainty in the {\it direction} of the
  ${\bf v}_A$ vector \citep{BL08}, give the same estimate in the case of {\it balanced} turbulence
  considered in \citet{GS95}.  In the {\it imbalanced} case it could lead to a modified relation.} $\tau_{casc}\omega>1$,
therefore the combination of this lower bound, that limits the strength of turbulence $\xi$ and the
tendency of turbulence to become stronger will make it ``critically balanced'' with $\xi\sim
1$. This critically balanced cascade is strong in a sense that cascading timescale is always of the
order of the dynamic timescale and will, therefore, have a Kolmogorov -5/3 spectrum. These critically balanced
perturbations will be strongly anisotropic with respect to the local mean magnetic
field. Using $\xi\sim 1$ and $w\sim k_\perp^{-1/3}$ one obtains $k_\|\sim k_\perp^{2/3}$, i.e. the
anisotropy will increase towards small scales without limit. One can further simplify Eq. 1 by
neglecting the term $(\delta w^\mp_\| \nabla_\|) \delta w^\pm$ which is much smaller than the mean
field term $(v_A \nabla_\|) \delta w^\pm$. After this Eq.~\ref{mhd} splits into two equations, one for
$\delta w^\pm_\|$, which, in this strongly anisotropic case, $k_\|\ll k_\perp$ represents slow (or
pseudo-Alfv\'en) mode and the equation for $\delta w^\pm_\perp$ which represent Alfv\'enic mode. The
equation for slow mode is passive and does not provide any back-reaction for the Alfv\'enic equation
which could be written in the following form:

\begin{equation}
\partial_t{\bf \delta w^\pm_\perp}\mp({\bf v_A}\cdot\nabla_\|){\delta \bf w^\pm_\perp}+\hat S ({\delta \bf w^\mp_\perp}\cdot\nabla_\perp){\delta \bf w^\pm_\perp}=0,\label{rmhd}
\end{equation}

One can study this purely Alfv\'enic dynamics and assume that the omitted slow mode has
similar cascade and similar statistical properties.  Eq.~\ref{rmhd} is known as reduced MHD
approximation or RMHD, see, e.g., \citet{kadomtsev,strauss}. RMHD equations provide further support
towards Goldreich-Sridhar model, as it has a precise symmetry with respect to anisotropy and the
strength of the mean field. Indeed, as long as one increases $v_A$ and stretches the fields in the
parallel direction, decreasing $\nabla_\|$, by the same factor, Eq. 2 will be
unchanged. Furthermore, a Kolmogorov argument of universality of nonlinear dynamics at each scale,
which is based on a two-parametric scaling symmetry, could be amended it with a proper scaling for
the anisotropy:

\begin{equation}
{\bf w} \to {\bf w}A,\ \ \ \lambda \to \lambda B, \ \ \  t \to t B/A, \ \ \  \Lambda \to \Lambda B/A, \label{symmetry}
\end{equation}

where $\lambda$ is a perpendicular scale, $\Lambda$ is a parallel scale, $A$ and $B$ are arbitrary
parameters. Due to this precise symmetry one can hypothesize that strong Alfv\'enic turbulence has a
universal regime, utilizing the same argumentation as \citet{kolm}. In nature, this universal regime
can only be achieved as long as $\delta w^\pm \ll v_A$. In numerical simulations, we can directly
solve the reduced Eq.~\ref{rmhd} , which has precise symmetry already built in. From practical
viewpoint, the statistics from the full MHD simulation with $\delta w^\pm \sim 0.1 v_A$ is virtually
indistinguishable from RMHD statistics and even $\delta w^\pm \sim v_A$ are fairly similar to the
former \citep{BL09a}. Note, that as we see above, both MHD and RMHD dynamics are essentially
three-dimensional. In this paper we use both full MHD simulations and RMHD simulations.
Statistically isotropic MHD simulation is used to determine a fraction of total energy contained in
the slow mode, while RMHD simulations are used to study properties of the universal
Alfv\'enic turbulence.

Previous numerical work confirmed scale-dependent anisotropy of the strong MHD turbulence
\citep[see, e.g.,][]{cho2000, maron2001}. The precise value of the spectral slope, however, was a
matter of debate. In particular, \citep{muller2005} claimed that the mean field strong turbulence
has a slope of $-3/2$.  This motivated adjustments to the Goldreich-Sridhar model
\citep{galtier2005,boldyrev2005,gogoberidze}. A model with so called ``dynamic alignment''
\citep{boldyrev2005, boldyrev2006} became popular after the alignment was discovered in numerical
simulations \citep{BL06}. Boldyrev model assumes that the alignment between velocity and magnetic
perturbations decreases the strength of the interaction, also it assumes that the alignment is a
power-law function of scale, increasing indefinitely towards small scales, modifying the spectral
slope of MHD turbulence from the $-5/3$ Kolmogorov slope to $-3/2$ slope. In this paper we debate
the assumption that the alignment is a power-law function of scale. We also prove that earlier
measurements of the slope were premature and were unable to reach asymptotic slope due to diffuse
locality of MHD turbulence. 

\begin{table}[t]
\begin{center}
\caption{Three-dimensional simulations}
  \begin{tabular}{c c c c c c}
    \hline\hline
Run  & Type & $n_x\cdot n_y \cdot n_z$ & Dissipation & $\langle\epsilon\rangle$ &  $L/\eta$ \\

   \hline

H1 & hydro &  $512^3$   &  $-3.02\cdot10^{-4}k^2$                & 0.091 &  190\\

H2 & hydro &  $1024^3$  &  $-1.20\cdot10^{-4}k^2$               & 0.091 &  370 \\

M1 & MHD &   $1024^3$  &  $-1.63\cdot10^{-9}k^4$                & 0.159 &  280 \\

R1 & RMHD & $256\cdot 768^2$ & $-6.82\cdot10^{-14}k^6$    & 0.073 &   280 \\

R2 &  RMHD & $512\cdot 1536^2$ & $-1.51\cdot10^{-15}k^6$  & 0.073  &  570 \\

R2.5 &  RMHD & $1536^3$ & $-1.51\cdot10^{-15}k^6$             & 0.073 &   570 \\

R3 & RMHD & $1024\cdot 3072^2$ & $-3.33\cdot10^{-17}k^6$ & 0.073 &  1100\\

   \hline

\end{tabular}
  \label{experiments}
\end{center}
\end{table}

\section{Numerical methods}
We used a pseudospectral dealiased code that was able to solve hydrodynamic, MHD and RMHD
equations. The RHS of Eq. 2 has an explicit dissipation term $-\nu_n(-\nabla^2)^{n/2}{\bf w^\pm}$
and forcing term ${\bf f}$. The code and the choice for numerical resolution, driving, etc, 
was described in great detail in our earlier publications \citep{BL09a, BL09b, BL10}.
Table 1 shows the parameters of the simulations. The Kolmogorov scale is defined as $\eta=(\nu_n^3/\epsilon)^{1/(3n-2)}$,
the integral scale $L=3\pi/4E\int_0^\infty k^{-1}E(k)\,dk$ (which was approximately 0.79 for R1-3).
Dimensionless ratio $L/\eta$ could serve as a ``length of the spectrum'', although usually spectrum
is around an order of magnitude shorter. 

\begin{figure}
\includegraphics[width=1.0\columnwidth]{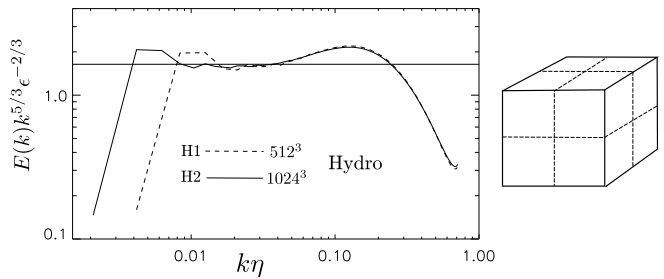}
\caption{Our hydrodynamic simulations reproduce $C_K=1.64$. As long as the turbulence is local and
  the effects of large scales could be neglected in the inertial range, the larger $1024$ cube could
  be seen as consisting of a eight smaller $512$ cubes, as on the right.}
\label{hydro}
\end{figure}

The resolution in the direction parallel to the mean magnetic field, $n_x$ was reduced by a factor of 3
for simulations R1-3. This was possible due to an empirically known lack of energy in the parallel
direction in $k$-space. We ran a simulation R2.5 which has full resolution in $n_x$ to compare with R2
and check the influence of this resolution reduction on the power spectrum. Although the bottleneck
effect was slightly less pronounced in R2.5 compared to R2, there was only a small influence in the
inertial range. We concluded that using $n_x$ reduced by a factor of 2 or 3 is possible.

For the purpose of this paper we used driving that had a constant energy
injection rate. In RMHD simulations R1-3 we drove turbulence to the amplitude that it will be strong
on the outer scale. R1-3 were started from lower-resolution simulation that reached stationary state
and were further evolved in high resolution for approximately 12 Alfv\'enic times, which, for strong MHD turbulence
also correspond to about 12 dynamical times. The averaged quantities were obtained for the last 6 Alfv\'enic times.
In all magnetic simulations M1, R1-3, we were using hyperviscosity ($n>2$) instead of normal viscosity. This
is possible due to the fact that bottleneck effect is much less pronounced in the MHD case, compared to hydro.

\section{Spectra and universality}
Much of the study of hydrodynamic turbulence was dedicated to Kolmogorov model which assumes a
universal cascade of energy through scales \citep{kolm}.  This model predicts that the power
spectrum of turbulence, $E(k)$, will be a power-law function of scale,

\begin{equation}
E(k)=C_K \epsilon^{2/3} k^{-5/3}.
\end{equation}

where $C_K$ is a Kolmogorov constant. It is well-known that this scaling is not precisely correct and
typically has an intermittency correction $(kL)^\alpha$, where $L$ is an outer scale
and $\alpha$ is a small number, around $0.035$ \citep{she}. However, in simulations or measurements
with small inertial range this correction can often be neglected. In particular, a compilation of
experimental results for hydrodynamic turbulence \citep{sreenivasan} suggests that a Kolmogorov
constant is universal for a wide variety of flows. High-resolution numerical simulations of
isotropic incompressible hydrodynamic turbulence \citep{gotoh} suggest the same value for the
Kolmogorov constant.

\begin{figure}
\includegraphics[width=1.0\columnwidth]{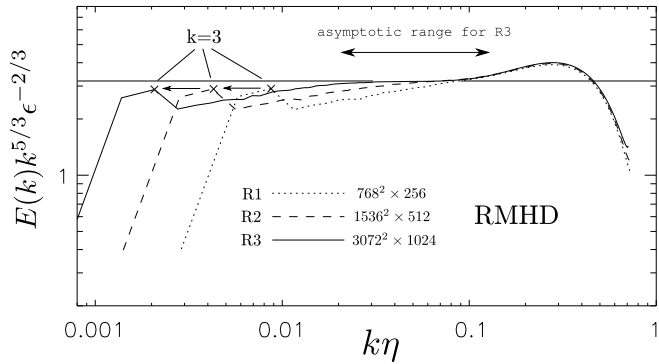}
\caption{Resolution study for MHD simulations. As MHD is less local than hydro, convergence require
  higher resolution.
  The estimate of Kolmogorov constant from averaged spectra for this purely
  Alfv\'enic turbulence is $C_{KA}=3.2\pm 0.2$.}
\label{mhd_ck}
\end{figure}

A robust method for determining the spectral slope and the Kolmogorov constant from simulations is a
resolution study \citep[see, e.g.,][]{gotoh}, when a number of numerical experiments are performed
with different resolution and the spectra are plotted with respect to the dimensionless wavevector,
$k\eta$.  A physical meaning of such a comparison is based on an assumption that a simulation with
higher numerical resolution can be considered both as a simulation resolving smaller physical scales
and as a simulation of a larger volume of turbulence (see Fig. \ref{hydro}).  This assumption is
true as long as turbulence can be considered local, i.e. the effects of driving can be neglected in
the inertial range.  Our hydrodynamic simulations reveal a good convergence of spectra with
numerical resolution and show a universal Kolmogorov constant consistent with the one obtained in
\citet{gotoh}.  Also the shape of the dissipation range is similar to the one in aforementioned
paper, showing a typical ``bump'' due to a bottleneck effect.  Despite moderate resolution, the
inertial ranges converge, which is due to locality of hydrodynamic cascade in spacial scales, making
it possible to consider higher and lower resolution simulations on a common ground, neglecting the
influence of large scales, where energy is provided by driving.

Fig.~\ref{mhd_ck} presents a resolution study for simulations R1-3 determining the spectral slope
and Kolmogorov constant for Alfv\'enic turbulence.  If the spectrum $-3/2$ was universal, the outer
scale point, corresponding to $k=3$ which is marked by a cross will go down from R1 to R3 by a
factor of around 1.26, instead it stays at about the same level, indicating that deviations from
$-5/3$ slope are small (note that the outer-scale point moves horizontally in Fig.~\ref{hydro} as
well). The flat part of Fig.~\ref{mhd_ck} in R3 simulation between $k=54\ (k\eta\approx 0.037)$ and
$k=91\ (k\eta\approx 0.063)$ with central frequency $k=70$ was fit to obtain Kolmogorov constant.
The value obtained in this fit was $C_K=3.2\pm 0.2$ where the error was mostly due to fluctuation
of spectrum in time.

\section{Dynamic alignment}
\begin{figure}
\begin{center}
\includegraphics[width=0.8\columnwidth]{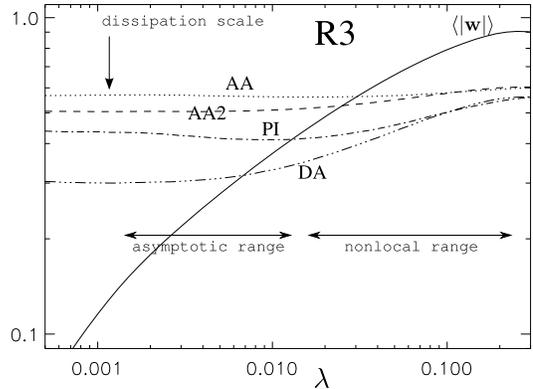}
\end{center}
\caption{Dynamic alignment seems to saturate towards small scales {\it before} the dissipation scale. This indicates that the scale-dependent alignment is a transient effect that is
present in simulations due to MHD turbulence being less local. The asymptotic regime of MHD turbulence is showing constant alignment, which will not modify the spectral slope $-5/3$
of strong MHD turbulence (GS95 slope).}
\label{align}
\end{figure}

It was suggested that the spectral slope of MHD turbulence is modified by so-called ``dynamic
alignment'' that increases indefinitely towards small scales.  Although the tentative correspondence
with theoretical scaling from Boldyrev model has been obtained with only one particular measure of
alignment, this was interpreted by some studies as a confirmation of the aforementioned model. In this paper we refer
to our earlier studies \citet{BL06,BL09b} that measured several types of alignment and their
dependence on scale. In these studies there were no conclusive evidence that all alignment measures
follow the same scaling. In this paper we confirm this finding with higher-resolution simulations, in addition
we found evidence that all alignment measures saturate, i.e. approach an asymptotic constant value
on small scales.  Fig. \ref{align} shows the alignment measures in R3, where AA, AA2, DA and PI are different
alignment measures: 
$AA=\langle|\delta {\bf w}^+_\lambda\times \delta {\bf w}^-_\lambda|/|\delta {\bf w}^+_\lambda||\delta {\bf w}^-_\lambda|\rangle$,
$AA2=\langle|\delta {\bf v}^+_\lambda\times \delta {\bf b}^-_\lambda|/|\delta {\bf v}^+_\lambda||\delta {\bf b}^-_\lambda|\rangle$,
$PI=\langle|\delta {\bf w}^+_\lambda\times \delta {\bf w}^-_\lambda|\rangle/\langle|\delta {\bf w}^+_\lambda||\delta {\bf w}^-_\lambda|\rangle$,
$DA=\langle|\delta {\bf v}^+_\lambda\times \delta {\bf b}^-_\lambda|\rangle/\langle|\delta {\bf v}^+_\lambda||\delta {\bf b}^-_\lambda|\rangle$, for more details and motivation, see \citet{BL06,BL09b}.
Having an inertial range of
around two orders of magnitude in scale, if alignment was proportional to $\lambda^{1/4}$ as in
\citet{boldyrev2005,boldyrev2006}, we would expect alignment increase by a
factor of $3.2$, while in reality the polarization intermittency PI increases only by a factor of
1.3, and dynamic alignment DA by a factor of 1.8. This is consistent with the range of $\lambda$
between $3$ and $10$.

We are not aware of any convincing physical argumentation explaining why alignment
should be a power-law of scale. \citet{boldyrev2006} provides an explanation arguing that alignment will
tend to increase indefinitely, but will be bounded by field wandering, i.e. the alignment on each scale
will be created independently of other scales (hence the term ``dynamic alignment'') and will be proportional
to the relative perturbation amplitude $\delta B/B$. But this directly violates precise symmetry of Eq.~\ref{rmhd}, i.e.
Eq.~\ref{symmetry} which states that nothing should depend on $\delta B/B$ as long as other quantities are scaled properly.
Physically, this means that field wandering can not destroy alignment or imbalance. Indeed a perfectly aligned state, e.g.,
with $\delta {\bf w}^-=0$ is a precise solution of Eq.~\ref{rmhd} and it is not destroyed by its own field wandering.
Additionally, \citet{BL09b} measured alignment in simulations of strong MHD turbulence
with different values of $\delta B_L/B_0$ and found very little or no dependence on this parameter.
Fig.~\ref{align} also compares alignment measure with a first-order structure function of the perturbation
amplitude, i.e. $\langle|\delta {\bf w}_\lambda|\rangle$. According to \citet{boldyrev2006} they should scale the same
way, but this is not observed.

To summarize, our numerical data are consistent with alignment measures becoming constant in the
inertial range and inconsistent with the hypothesis that they depend as $\lambda^{1/4}$ on scale.
This finding is important, because if alignment is constant on scale
in the asymptotic regime, there is no reason to expect that the power-law scaling of turbulence will
deviate from its $-5/3$ value for strong Goldreich-Sridhar turbulence. This result is further
supported by the results of the previous section where a steeper asymptotic spectra has been
observed.

\section{The amount of slow mode and the total Kolmogorov constant for MHD turbulence}
Full incompressible MHD turbulence have a cascade of slow mode, which was not included in our
reduced MHD simulations R1-3. Although in nature slow mode is often damped, it is normally present
in full MHD incompressible simulations, e.g. the ones presented in \citet{biskamp}. The passive
cascade of slow mode will have the same energy spectral slope as a Alfv\'enic mode, and, assuming that
the ratio of slow to Alfv\'enic energies is $C_s$, the {\it total} Kolmogorov constant for MHD
turbulence will be expressed as

\begin{equation}
C_K=C_{KA}(1+C_s)^{1/3}.\label{ck}
\end{equation}

The ratio $C_s$ is supposedly depend on how the MHD turbulence is driven. However, historically, previous
studies simulated MHD turbulence with zero mean field, either decaying or driven with statistically isotropic forcing,
e.g. \citet{muller2005}. In this idealized case Kolmogorov constant
has been measured, although with fairly limited resolution \citep{biskamp}. In this paper we will use a less
straightforward approach, by measuring $C_s$ from a simulation with zero mean field and
substituting it into Eq.~\ref{ck}. This approach is motivated by our finding that MHD turbulence is less
local and therefore it is much harder to achieve an asymptotic universal cascade if one uses
zero-mean field simulation. Indeed, one has to observe the transition to the strong local mean field
case, which will require at least a couple of order of magnitude in scale and subsequently a
transition to universal cascade, which as we observed in previous section, takes about two orders of
magnitude in scale, as long as the power-law scaling and Kolmogorov constants are concerned. It is,
therefore, impossible to directly measure the properties of universal cascade in zero mean field simulations
of currently available simulations. The ``natural'' value of $C_s$ is unity, because the
incompressible MHD equations have four degrees of freedom, out of which Alfv\'enic mode uses two and
slow mode also uses two. Having the same amount of degrees of freedom and the isotropic driving that
does not prefer any direction we would expect that the energy will be distributed equally between
the modes. We measured how energy is partitioned on small scales of simulation M1 by making a local
Fourier transform of smaller cubes and decomposing into modes with respect to the local mean
field. The actual partition of energy shows $C_s$ being around $1.3$. Although statistical errors in
this measurement are small, it is hard to claim a particular value of $C_s$ based on a numerical
simulation with a finite resolution. Conservatively, we will assume that $C_s$ is between $1$, which
is equipartition, and $1.3$, which is observed in our simulation M1. The total Kolmogorov constant
will be estimated as $4.1\pm0.3$.

\section{Scale locality and Kolmogorov constant}

The energy flux through scales in both MHD and hydrodynamic turbulence can be expressed as a certain
third order {\it signed} structure function
divided by scale and has to be scale-local due to an upper analytical bound on contributions from different $k$ wavebands
\citep[see, e.g.][]{aluie2010}. This upper bound, however, is well applicable to similar third order {\it unsigned}
structure function. This {\it unsigned} third order structure function is related by self-similarity
hypothesis to second order structure function, which is a measure of energy. Therefore, we would expect
that the ratio of {\it unsigned} third order structure function to the signed one will scale approximately
as $C_K^{3/2}$. This seriously limits the bound on scale locality from practical standpoint
as long as $C_K$ becomes large, i.e., the energy transfer becomes less efficient \citep{BL10}.
Indeed, if the define ``scale locality'' as a ratio of largest to smallest wavevectors $k_1/k_2$ which still
significantly contribute to energy flux through some central wavevector $k_0$, then this ratio will have an upper
bound that scale asymptotically with Kolmogorov constant as $C_K^{9/4}$. In other words, inefficient energy
transfer can still be very local, but it is also possible that it is less local than efficient energy transfer.
A nonlocal or diffuse energy transfer {\it must} be inefficient and {\it must} have a high value
of Kolmogorov constant.

A comparative study of energy spectra in MHD and hydro turbulence in \citet{BL09b}
revealed that the bottleneck effect is less pronounced or altogether absent in MHD simulations, while in
hydro it is always present, both in simulations with normal ($n=2$) and hyperviscosity ($n>2$).
This was interpreted as an indication that MHD cascade is less local. Now, our measurement of Kolmogorov constant
revealed that MHD energy transfer is less efficient, therefore MHD cascade may be less local than hydro cascade.
In view of all numerical evidence available today, MHD cascade is most likely less local than hydro cascade. 

\section{Discussion}
Previous measurements of the slope usually relied on the highest-resolution simulation and fitted the slope
in the fixed $k$-range close to driving scale typically between $k=5$ and $k=20$. In this paper we argue that such a fit is
unphysical and instead one should fit a fixed $k\eta$ range. In the former case the result would be a shallower spectral
slope due to proximity to the outer scale and driving. In the latter
case the effect of the driving will diminish with increasing resolution and one will observe shallower spectra at small
resolutions that will become steeper with increasing resolution.


Earlier measurements of Kolmogorov constant in MHD turbulence reported lower values than this study,
e.g. $C_K=2.2$ in \citet{biskamp}.  We believe this is due to insufficient resolution in those
simulations, which prevented the observation of the asymptotic regime.  In particular, in the case of
statistically isotropic simulations like the ones in \citet{biskamp} a transition to
small scale subAlfv\'enic regime precede the transition to asymptotic regime. These two transitions
require numerical resolution that is even higher than the highest resolution presented in this paper and for now seems
computationally impossible. Our own statistically isotropic simulation M1 shows Kolmogorov constant of 3.5,
which is still only a lower limit, consistent with 4.1 derived in this paper.
For M1 and similar lower-resolution simulations the estimate of $C_K$ continues to grow with
increasing resolution, which supports argumentation above.


In this paper we treated so called balanced case, where the rms amplitudes of the $w^\pm$ were
statistically the same. A number of attempts to generalize the GS95 model has been made recently
\citep{lithwick2007,BL08,chandran2008, PB09}. Some of these models can be rejected by numerics, in
particular the model based on alignment \citep{PB09} is grossly inconsistent with the dissipation
rates measured in imbalanced numerical simulations \citep{BL09a,BL10}.

\begin{acknowledgments}
LA-UR 10-07274.
This research was supported in part by the NSF through
TeraGrid resources provided by TACC under grant number TG-AST080005N.
\end{acknowledgments}

\ \ \

\end{document}